\documentclass[aps,pra,reprint,twocolumn]{revtex4-2}

\usepackage{graphicx}
\usepackage{amsmath}
\usepackage{float}
\usepackage{hyperref}
\usepackage{orcidlink} 
\usepackage{upgreek}
\usepackage{gensymb}
\hypersetup{
    colorlinks=true,
    linkcolor=blue,
    citecolor=blue,
    urlcolor=blue
}
\def\americanIsolect{1}
\usepackage{./isolects}

\begin{document}
	
\title{Polar\IS{}ation-sensitive vector magnetometry using nitrogen-vacancy cent\RE{}s in diamond}
\author{M.S.J. Barson\,\orcidlink{0000-0003-0247-5619}\textsuperscript{\textdagger}}
\email{michael.barson@monash.edu}

\affiliation{School of Physics and Astronomy, Monash University, Victoria, 3800, Australia}

\author{T.J. Christie\,\orcidlink{0009-0003-3697-5037}}
\thanks{These authors contributed equally to this work.}
\affiliation{School of Physics and Astronomy, Monash University, Victoria, 3800, Australia}

\author{J.P. Duff\,\orcidlink{0009-0008-2644-7528}}
\thanks{These authors contributed equally to this work.}
\affiliation{School of Physics and Astronomy, Monash University, Victoria, 3800, Australia}

\author{K. Helmerson\,\orcidlink{0000-0002-6387-0949}}
\affiliation{School of Physics and Astronomy, Monash University, Victoria, 3800, Australia}

\begin{abstract}
The phase relationships between the vector components of a time-varying field define its polar\IS{}ation, which can provide insight into the structure of a magnetic field source and its environment.
The goal of this work is to measure magnetic field elliptical polar\IS{}ation parameters, and demonstrate their utility through direct observation of field source rotations and environmental coupling.	
An ensemble of nitrogen-vacancy (NV) cent\RE{}s in diamond is used to perform phase-sensitive ac vector magnetometry, enabling direct measurement of field polar\IS{}ation ellipses using complex spin-resonance solutions to the NV Hamiltonian. 
Measurements of wire-coil sources demonstrated retrieval of linear and elliptical magnetic polar\IS{}ations, allowing direct observation of source rotation and mutual coupling between coils.
These results establish magnetic polarimetry as a practical extension of ac vector magnetometry, and promise applications in vector ac susceptometry for anisotropic materials character\IS{}ation, and the analysis of reactive electrical systems.  
\end{abstract}

\maketitle
\section{INTRODUCTION}
AC magnetometry is a widely used technique for probing magnetic environments, being deployed within condensed matter physics \cite{casolaProbingCondensedMatter2018, schirhaglNitrogenVacancyCentersDiamond2014}, materials science \cite{gongDiscoveryIntrinsicFerromagnetism2017}, and electrical circuit analysis \cite{sengottuvelMicrowavefreeImagingMagnetometry2025}. 
Such magnetometry techniques are well established using many different sensor technologies \cite{lenzMagneticSensorsTheir2006} (SQUIDs \cite{jenksSQUIDsNondestructiveEvaluation1997}, atomic sensors \cite{marasliDipoleLocalizationUsing2026}, Hall effect sensors \cite{popekLocalizationMethodMagnetic2013}, magnetoresistive sensors \cite{chibaThreedimensionalLocalizationRotating2020}, fluxgates \cite{yangNewCompactFluxgate2014}, \textit{etc.}) in a mix of scalar and vector measurement configurations.
Those vector magnetometers, however, seldom make direct use of the phase information implicitly captured in their vector reconstruction.

While measuring the global phase of an isolated ac field is not generally useful without a reference phase, in the context of ac susceptometry and ac vector magnetometry, meaningful information is conveyed in the relative phase changes of the field. 
AC susceptometry measures the complex magnetic susceptibility (dissipative and reactive) of materials via phase and amplitude measurements of a magnetic field induced in that material by a reference field.
Classic ac susceptometry is typically performed on bulk isotropic materials, such as superconductors \cite{gomoryCharacterizationHightemperatureSuperconductors1997}, spin glasses \cite{mydoshSpinGlassesRedux2015}, or molecular nano-magnets \cite{gatteschiMolecularNanomagnets2006}, using excitation coils and pickup coils to measure the induced field \cite{nikoloSuperconductivityGuideAlternating1995}.
Whilst this is highly effective for measuring bulk material responses, directional and spatial information about the induced field is generally lost.
Similarly, in ac vector magnetometry, the relative phases of magnetic vector field components in 3D space define the field polar\IS{}ation.
This polar\IS{}ation contains information about the source's geometry and internal structure, alongside any reactive or anisotropic interactions occurring within the environment.
One could combine ac susceptometry and vector magnetometry to measure or image the polar\IS{}ation of the field induced in a material by a reference field, determining the complex, directional susceptibility of \textit{anisotropic} materials, or alternatively, the reactance of structured electrical circuits. 

\textit{Vector} ac magnetometry has been demonstrated using a variety of measurement platforms, both classical and quantum. 
Many such demonstrations util\IS{}e projective magnetometers relying on mechanically aligned orthogonal sensors, which introduce alignment errors to the vector reconstruction \cite{chibaThreedimensionalLocalizationRotating2020, elgeeCompleteThreedimensionalVector2024}.
Magnetometers based on ensembles of nitrogen vacancy (NV) cent\RE{}s in diamond have the advantage of having four crystallographically-fixed NV orientations in the diamond lattice, allowing for vector measurements without alignment error \cite{steinertHighSensitivityMagnetic2010, wangHighdynamicrangeVectorMagnetometry2026}.
The NV cent\RE{} in diamond is an established quantum sensor for magnetometry \cite{taylorHighsensitivityDiamondMagnetometer2008}, offering robust room-temperature operation, high sensitivity, and intrinsic vector readout capabilities \cite{dohertyNitrogenvacancyColourCentre2013, degenQuantumSensing2017}.
Such NV-based magnetometry has been demonstrated at a range of frequencies from dc \cite{barrySensitiveAcDc2024, lonardLimitsAbsoluteVector2026} and linewidth-limited MHz \cite{barrySensitiveAcDc2024, zhengMicrowaveFreeVectorMagnetometry2020, broadwayAnticrossingSpinDynamics2016}, through to hundreds of MHz and GHz using dynamical decoupling protocols \cite[Fig.~5c]{rondinMagnetometryNitrogenvacancyDefects2014, ranaImagingMicrowaveMagnetic2025} and heterodyne protocols \cite{wangPicoteslaMagnetometryMicrowave2022a, taylorHighBandwidthHeterodyne2026}.
These measurements have been performed in a variety of readout geometries, including single NV readout \cite{mazeNanoscaleMagneticSensing2008}, ensembles \cite{barrySensitiveAcDc2024, childressBiasFieldFreeOperationNitrogenVacancy2025, kumarHighDynamicrangePortable2024}, widefield \cite{scholtenWidefieldQuantumMicroscopy2021, bassoWidefieldMicrowaveMagnetic2024, rebeirroMicrowavefreeWidefieldMagnetometry2023, horsleyMicrowaveDeviceCharacterization2018}, and scanning probe \cite{choiScanningMiniaturizedMagnetometer2025, heHighprecisionThreedimensionMicrowave2026, troiseHighspeedMagneticfieldImaging2026} configurations.
While phase-sensitive detection is routinely employed in NV ac magnetometry, the relative phase relationships between vector field components are rarely treated as a useful measured quantity.
Consequently, magnetic field polar\IS{}ation has remained largely unexplored as a practical observable for magnetic field character\IS{}ation and source/environment analysis. This gap is the focus of the present manuscript.

\section{PHASE-SENSITIVE VECTOR MAGNETOMETRY} 
Time-varying vector fields can have a unique phase delay associated with each spatial component.
These phases capture information about the polar\IS{}ation of the field, or the structure of the field source and its environment.
For example, these phases could be the response of some lossy or anisotropic material or the reactance of electrical currents.
Commonly, these phases are described using phasor notation, where the real-valued magnetic field $\mathcal{B}(\mathbf{r},t) = \Re\left[\mathbf{B}(\mathbf{r})e^{i\omega t} \right]$ has the non-time-varying part described by the complex-valued phasor

\begin{align}
	\begin{aligned}
		\mathbf{B}(\mathbf{r}) = \left (\begin{array}{c}
			B_x(\mathbf{r})e^{i\phi_x(\mathbf{r})} \\ B_y(\mathbf{r})e^{i\phi_y(\mathbf{r})} \\
			B_z(\mathbf{r})e^{i\phi_z(\mathbf{r})}
		\end{array} \right ).
	\end{aligned}
\end{align}

This work shows how this phasor can be directly measured by coherently observing the Zeeman shift from an ensemble of NV cent\RE{}s in diamond.

\begin{figure}
	\centering
	\includegraphics[width=0.48\textwidth]{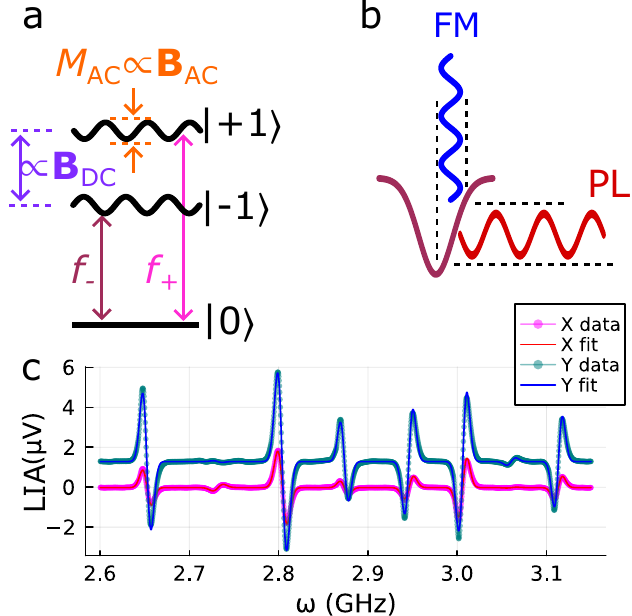}
	\caption{\textbf{(a)} Spin levels of the NV cent\RE{} undergoing dc and ac Zeeman shifts.
	\textbf{(b)} Depiction of how FM-modulated microwaves resonant on an ODMR spin-resonance (or equivalently an ac magnetic field) results in PL modulation.
	\textbf{(c)} ODMR spectra when demodulating at the $\mathbf{B}_\text{AC}$ frequency of 777 Hz.}
	\label{fig:levels-odmr}
\end{figure}

\section{EXPERIMENTAL METHOD} 
\subsection{Continuous-Wave Phase-Sensitive Detection}
As shown in Fig.~\ref{fig:levels-odmr}(a), when an ac magnetic field is applied, a time-varying Zeeman shift is induced.
Continuous-wave observation of such an ac magnetic field -- or equivalently, frequency modulation of the microwave carrier [Fig.~\ref{fig:levels-odmr}(b)] -- results in a modulation of the photoluminescence (PL).
By using phase-sensitive detection (here, a lock-in amplifier (LIA)) the response of each NV cent\RE{} orientation to the ac field can be measured.
Figure~\ref{fig:levels-odmr}(c) shows a typical optically detected magnetic resonance (ODMR) spectrum when demodulating at the ac magnetic field frequency (or FM modulation frequency).
These spectra have a derivative-like lineshape, due to the LIA measuring the first-order changes in fluorescence from the FM/ac modulations.
With a dual LIA, the in- and out-of-phase quadrature components of the signal ($X$ and $Y$) can be measured simultaneously.

\subsection{Calibration and Reconstruction Procedure}
Fitting independent Gaussian-derivative lineshapes to each quadrature, for each resonance, yields a complex amplitude $A = A_X + iA_Y$ for each resonance.
In the controlled FM modulation case, we set the exact frequency deviation $M_\text{FM}$ such that microwave magnetic field $B_{MW} \propto \cos\left(\omega t + M_\text{FM}\cos\omega_\text{FM}t\right)$, where $\omega$ is the applied microwave frequency.
We can use this to calibrate the optical response of each spin resonance to a particular frequency modulation.
The complex modulation of each spin-resonance under an unknown ac magnetic field can then be found as $M_\text{AC} = A_\text{AC} \times M_\text{FM}/\left | A_\text{FM} \right | $. 
Here $A_\text{FM}$ and $A_\text{AC}$ are the complex amplitudes from the quadrature lineshapes due to the FM and ac magnetic fields, respectively, but now in units of spin-resonance frequency shift (\textit{e.g.}, MHz). 
This calibration is only valid for small (approximately linear) shifts to the spin-resonance, where the modulation depth is much smaller than the spin-resonance linewidth.

To use these modulations for determining an ac magnetic field, we now use our knowledge of the spin-Hamiltonian.
The spin-Hamiltonian for a simple Zeeman shift is well known \cite{loubserOpticalSpinpolarisationTriplet}, $H=D\left(S_z^2-2/3\right)+\gamma\,\mathbf{B}\cdot\mathbf{S}$, where $D=2.87$ GHz is the zero-field splitting, $\gamma\approx28$ GHz/T is the gyromagnetic ratio, $\mathbf{B}$ is the magnetic field vector and $\mathbf{S}$ is a vector of the ($S=1$) spin operators.
The eigenvalues of the spin-Hamiltonian result in two magnetic field-dependent spin-resonance frequencies $f_+(\mathbf{B})$ and $f_-(\mathbf{B})$ between the $\left | 0 \right >$ and $\left | \pm 1 \right >$ spin-levels.
The ac component of the magnetic field creates a peak-to-peak frequency modulation of $M_\pm(\mathbf{B}) = f_\pm(\mathbf{B}_\text{DC} + \mathbf{B}_\text{AC}) - f_\pm(\mathbf{B}_\text{DC} - \mathbf{B}_\text{AC})$ for each spin-resonance. 
To incorporate phase into the spin-resonance, we now use a complex description of the magnetic field (i.e., $\mathbf{B} = \mathbf{B}' + i \mathbf{B}''$), yielding a complex modulation for each spin-resonance. This is determined by solving the Hamiltonian eigenvalues for the real and imaginary field components separately (i.e. 
$M_\pm(\Re[\mathbf{B}]) + i M_\pm(\Im[\mathbf{B}])$). Here the complex nature of the system is unrelated to the quantum-mechanical phase of the spin (\textit{e.g.} $\left[S_x,S_y\right] = iS_z$); it is just a convenient way to define the in- and out-of-phase response of the system to the ac magnetic field with a single complex quantity.

While phase information is inherently accessible via phase sensitive detection (like dual-phase lock-in demodulation), its rigorous translation into a full 3D vector trajectory is non-trivial due to systematic spatial alignment vulnerabilities \cite{elgeeCompleteThreedimensionalVector2024}.
A key metrological advantage of using the NV ensemble for this polarimetric reconstruction is the intrinsic reference frame provided by the diamond host crystal. 
Unlike classical inductive coils or Hall probes, where subtle mechanical misalignment errors introduce artificial phase shifts between spatial components, the rigid, crystallographic orientation of the four distinct NV sub-ensembles yields an over-determined coordinate basis \cite{maertzVectorMagneticField2010}. 
This structural symmetry permits the precise elimination of instrumental or electronics-induced phase lags from the physical phase shifts generated by the target environment \cite{wangNanoscaleVectorAC2021}.

To solve the ensemble Hamiltonian for the resonant frequencies given an incident ac magnetic field, it is easiest to first transform the total field from the crystal/lab coordinates into each sub-ensemble's local coordinate system via the rotation operator ($\mathcal{R}_i$) such that $\mathbf{B}_{\text{NV}_i} = \mathcal{R}_i \mathbf{B}_\text{lab}$. 
The same transformation is applied independently to the real and imaginary components of the complex magnetic field.
Solving the real and imaginary Hamiltonians in each local coordinate system and taking only the ac components yields the two complex modulation values $M_\pm(\Re[\mathbf{B}]) + i M_\pm(\Im[\mathbf{B}])$. 
We now have eight separate complex modulation ($M$) values; both $M_+$ and $M_-$ spin resonances for all four separate NV sub-ensembles.

We can now optim\IS{}e a cost function comparing the measured complex resonances from our NV ensemble to a simulated ensemble model which takes in a complex magnetic field as its argument.
Fortunately, this process is exactly the same as it is for real-valued fields.
We use least squares fitting and minim\IS e the cost function for both $\mathrm{m_s}$ spins (i) and each sub-ensemble orientation (k). 

\begin{align}
	\begin{aligned}
			 C(\mathbf{B}_\text{guess}) = & \sum_{m=\pm 1}\sum_{k=1}^4 \left[ M_{\text{data},m,k}\right. \\
			& \left. -M_{\text{model},m}(\mathcal{R}_k[\mathbf{B}_\text{guess}]) \right]^2.\label{eq:amplitude-fitting}
	  	\end{aligned}
\end{align} 

\section{EXPERIMENTAL DEMONSTRATIONS}
\begin{figure*}
	\centering
	\includegraphics[width=0.99\textwidth]{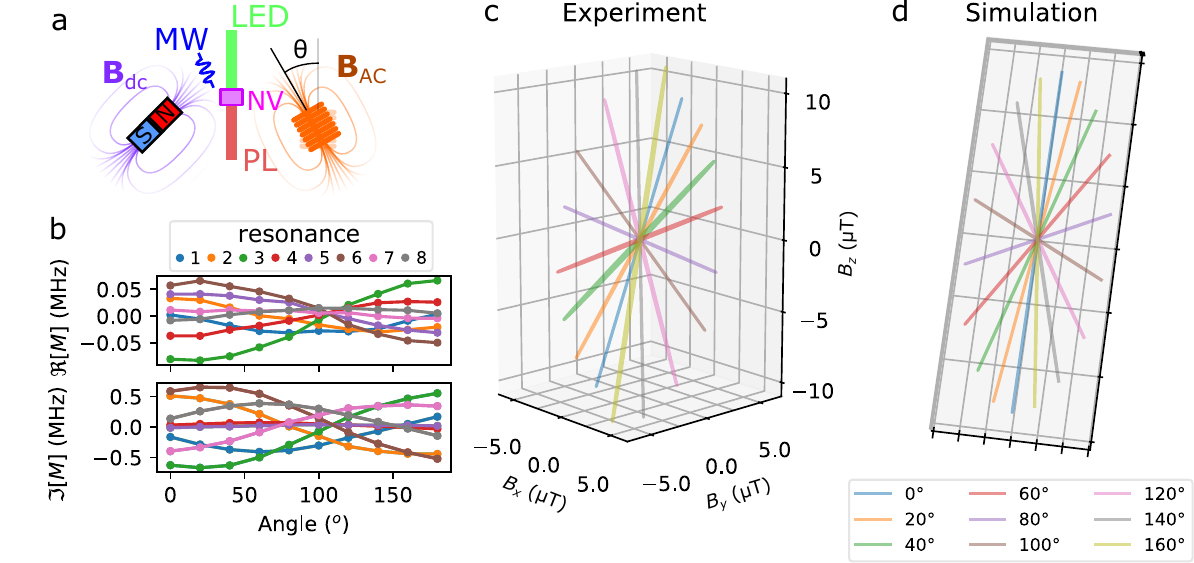}
	\caption{\textbf{(a)} Top-down view of rotating coil experiment, the coil was approximately 30 cm away from the diamond and rotated in the horizontal plane by $\theta$. Light pipes were used to couple the excitation and PL light in and out of the diamond. 
	\textbf{(b)} In- and out-of-phase modulations of the eight spin resonances in the NV ensemble. 
	\textbf{(c)}--\textbf{(d)} 3D polar\IS{}ation ellipses of the measured magnetic field and \textsc{COMSOL Multiphysics\textsuperscript{\textregistered}} model of the rotating coil experiment. The small ellipticity is indiscernible on these plots. 
	The \textsc{COMSOL\textsuperscript{\textregistered}} data shows no units, as it was only used to qualitatively compare the magnetic fields. 
	The experimental data is measured in the crystal coordinate system, which is arbitrarily positioned with respect to the laboratory coordinate system. 
	Similarly, the starting angle of the rotation of the coil was also arbitrary.}
	\label{fig:ellipses-rotating-coil}
\end{figure*}

We perform two separate measurements to demonstrate this technique.
First, to measure the vector sensing capability of this technique, we rotated a single magnetic coil nearby our diamond [Fig.~\ref{fig:ellipses-rotating-coil}(a)].
Second, to demonstrate the polar\IS{}ation sensing capability, we used two orthogonal crossed coils [Fig.~\ref{fig:ellipses-crossed-coils}(b)] and recorded the spectra for currents in both coils simultaneously and independently.
For each measurement, we also applied a constant dc bias-field from a nearby permanent magnet, which splits the spectra into eight separate resonances.

\subsection{Rotating Coil Demonstration}
An informative way to visual\IS{}e the resulting complex magnetic fields is through 3D polarization ellipses, analogous to those commonly used to represent the state of polarization of electromagnetic waves \cite[p.~7]{CollettObservablesPolarizedLight2005}.
These ellipses are parameter\IS{}ed as $\mathbf{p} = \Re \left[\mathbf{B}e^{i\phi} \right]$, where the phase parameter $\phi$ is swept through $0\le\phi<2\pi$.
These show the path that the magnetic field vector takes through one full cycle.
Figure \ref{fig:levels-odmr}(c) shows the ODMR response of the NV ensemble to a rotating coil, where there is a much larger out-of-phase ($Y$) component due to the $-\pi/2$ phase shift of the mostly inductive load of the coil. 
Figure \ref{fig:ellipses-rotating-coil}(c) shows the corresponding optim\IS{}ed magnetic fields from rotating the coil relative to the fixed diamond. 
Here we clearly see a rotating dipole pattern as the coil is rotated with some very slight ellipticity [eccentricity = 0.9983(7)], indicating orthogonal spatial components which are slightly out-of-phase with respect to each other.
In this measurement, the diamond was close ($\sim$ 5 cm) to a conductive alumin\IUM{} breadboard, which carries eddy currents induced by the applied ac field.
These current loops produce small out-of-phase magnetic fields which are detected by our measurement.
A simple \textsc{COMSOL\textsuperscript{\textregistered}} \cite{comsol64} model of a current-carrying coil above an alumin\IUM{} slab [Fig.~\ref{fig:ellipses-rotating-coil}(d)] shows agreement with this behavi\OU{}r.
We see reduced eccentricity for the experimental data than the \textsc{COMSOL\textsuperscript{\textregistered}} model [eccentricity = 0.9999(2)], which we attribute to the signal magnitude exceeding the linear regime of the spin-resonance, artificially attenuating the measured signal.
This could be alleviated by either examining the high-order harmonics \cite{barsonNanoscaleVectorElectric2021} of the applied signal ($2\omega_\text{AC}$, $3\omega_\text{AC}$, \textit{etc.}) or reducing the signal amplitude.

\subsection{Crossed-Coil Polarimetry}
\begin{figure}
	\centering
	\includegraphics[width=0.4\textwidth]{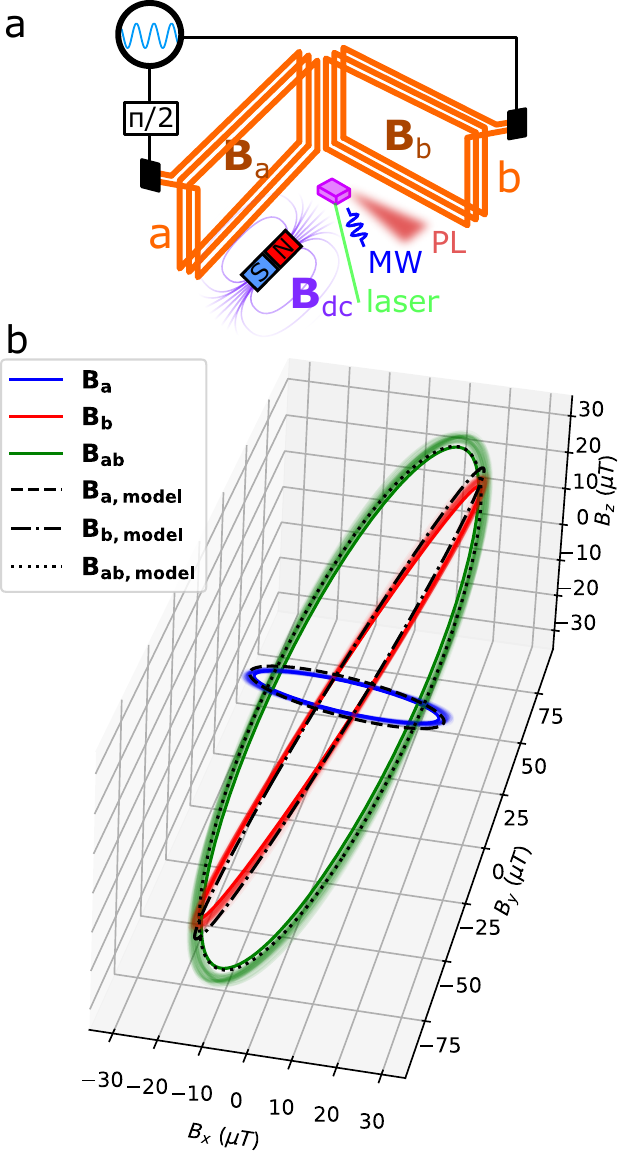}
	\caption{\textbf{(a)} Depiction of the crossed coil experiment, where two coils are placed orthogonal to each other. 
	\textbf{(b)} Polar\IS{}ation ellipses of the measured magnetic field from the orthogonal coil experiment. 
	The thickness of the lines signifies the experimental error. 
	The black lines show the model from equation (\ref{eqn:crossed-coils-model}).}
	\label{fig:ellipses-crossed-coils}
\end{figure}

For the case of the crossed coils (labe\LL{}ed $\mathrm{a}$ and $\mathrm{b}$ in Fig.~\ref{fig:ellipses-crossed-coils}(a)), it is expected that the individual coils produce linearly polar\IS{}ed magnetic fields $\mathbf{B}_\text{a}$ and $\mathbf{B}_\text{b}$ which combine with a $\pi/2$ phase shift to produce a circularly polar\IS{}ed field ($\mathbf{B}_\text{ab}$). However, it can be seen in Fig.~ \ref{fig:ellipses-crossed-coils}(b) that the $\mathbf{B}_\text{a}$ and $\mathbf{B}_\text{b}$ fields alone were elliptical in their polar\IS{}ation. 
For this measurement, the diamond was mounted on a printed circuit board (PCB) that was suspended midair by stainless steel mounts, which are not expected to have significant eddy currents. 
The ellipticity is caused by the mutual inductance of coils, where the magnetic field from one coil induces current in the other, introducing some out-of-phase components and ellipticity of the magnetic field. 
To determine this effect, we mode\LL{}ed the system as,
\begin{align}
	\begin{aligned}
		\mathbf{B}_\text{a} &= e^{i\alpha} \left|\mathbf{B}_\text{a}\right| \left(\mathbf{\hat{B}}_\text{a} + i m_c \mathbf{\hat{B}}_\text{b}\right) \\ 
		\mathbf{B}_\text{b} &= e^{i\beta} \left|\mathbf{B}_\text{b}\right| \left(\mathbf{\hat{B}}_\text{b} + i m_c \mathbf{\hat{B}}_\text{a}\right) \\
		\mathbf{B}_\text{ab} &= e^{i\kappa}\left( \mathbf{B}_\text{a} + \mathbf{B}_\text{b} \right),
		\label{eqn:crossed-coils-model}
	\end{aligned}
\end{align}
where $m_c$ is some mutual coupling parameter and $\alpha,\beta$ and $\kappa$ are some global phase factors to enable fitting. 
This model is plotted in Fig.~\ref{fig:ellipses-crossed-coils}(b) and shows good agreement with the experimental data, finding a mutual coupling factor of $m_c \approx 0.24$.

For both of the measurements, if we were unable to resolve and interpret the phase and the magnitude of the vector components, then we would have incorrectly measured the magnetic field vector.
Further improvements could be made by simply examining the higher harmonics of the spin-resonances, allowing for greater dynamic range; this would not change the complex magnetic field measurement method we have demonstrated.

\subsection{Measurement Uncertainty}
For both measurements, uncertainty was propagated using a Monte-Carlo bootstrap method \cite{efronIntroductionBootstrap1994}, where we generated new spectra which had random Gaussian noise added with the same standard deviation as the experimental data.
We then repeated the entire analysis process on the new spectra.
By examining the distribution of the semi-major and semi-minor ellipse axes lengths, we arrive at a geometric mean uncertainty of 7.7 nT and 0.6 $\upmu$T for the rotating coil and crossed coils respectively.
The significantly better uncertainty for the rotating coil was due to greater PL light collection using a light-pipe, whereas the crossed coil used a conventional microscopy objective.

\section{APPLICATIONS AND FUTURE DIRECTIONS}
\subsection{Coupled Electromagnetic Systems}
We have shown above that our protocol provides a useful vector probe of reactive (inductive or capacitive) current dynamics and field polar\IS{}ation, without requiring the high-frequency sampling and intensive data processing necessary for real-time baseband NV vector magnetometry. 

One application of this technique is in extending the crossed-coil experiment for the non-invasive character\IS{}ation of comparable coupled electromagnetic systems, including resonance circuits, inductors/transformers, and wireless power-transfer systems.
Conventional analysis of such systems is typically performed by impedance anal\YS{}ers or vector network anal\YS{}ers, which require probing access to terminal voltages and currents. 
These approaches model and return lumped-element quantities such as inductance, capacitance, coupling factor, and quality factor, but return no information about the geometry of electromagnetic sources.
In contrast, magnetic polarimetry can provide a direct view of the spatial geometry and coupling of multiple field sources with distinct field orientations and phase delays, without requiring invasive electrical access.
The orthogonal-coil experiment presented here represents a simple example of this method, but the same principle could be applied to more general systems. 
A potential example is a wireless power transfer system, where the combined field generated by transmitter and receiver coils may contain geometric information relating to relative coil alignment and coupling, thereby potentially providing a route for estimating and correcting position and orientation misalignment.

\subsection{AC Susceptometry of Anisotropic Materials}
\begin{figure*}
	\centering
	\includegraphics[width=0.99\textwidth]{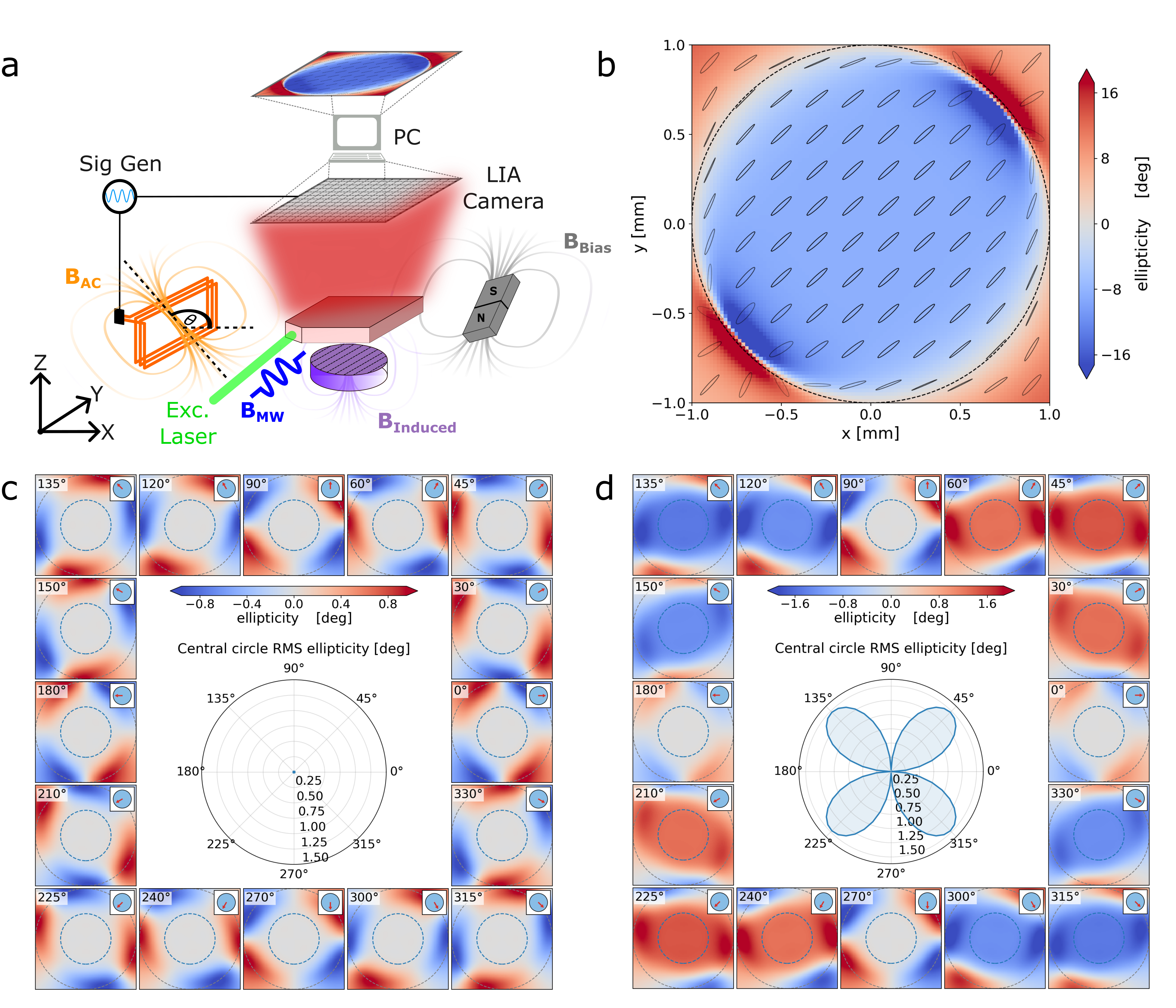}
	\caption{Simulated imaging of the frequency-dependent anisotropy in an MnZn ferrite. 
	(a) A 2 mm $\times$ 2 mm $\times$ 0.5 mm bulk NV ensemble diamond resting on the 1 mm-radius face of an anisotropic cylindrical ferrite experiences differing levels of field ellipticity as the ferrite is excited by a planar B-field from different directions. 
	The standard laser excitation, microwave addressing, and bias field are used for ODMR readout. 
	(b) An image of the pixel-wise ellipticity of the magnetic field induced in a ferrite material with exaggerated anisotropy ($\mu_x/\mu_y \approx 5.5$) to make ellipse overlay clearer (10 MHz, 45$^\circ$ excitation).
	(c), (d) The ellipticity in the total magnetic field (excitation + induced) experienced by an NV ensemble with a rotating excitation polar\IS{}ation angle. 
	The \GREY{} circle shows the ferrite outline, and the blue circle indicates the region considered in the central polar plot. 
	The isotropic sample (c) shows negligible measurable field ellipticity above the ferrite sample, whilst a representative MnZn ferrite sample with $\mu_x/\mu_y \approx 1.15$ anisotropy produced a total field with up to 1.5\degree \ of ellipticity.}

	\label{fig:polarimetry-imaging}
\end{figure*}

Beyond electrical circuits, this technique also creates a robust framework for magnetic ellipsometry, serving as a direct analog\UE{} to optical ellipsometry, where polar\IS{}ation changes are used to infer directional material properties.
When integrated with high-resolution widefield imaging using a lock-in camera (pixelwise lock-in amplification) \cite{scholtenWidefieldQuantumMicroscopy2021}, this could be deployed to examine highly local\IS{}ed directional, anisotropic, and absorptive magnetic features [Fig.~\ref{fig:polarimetry-imaging}(a)]. 
Existing ac susceptometry techniques are conducted by taking a scalar phase measurement, and are limited to monitoring bulk relaxation, demagnet\IS{}ation, or screening properties of a sample \cite{boothDeterminingACSusceptibility2021, dilleyACSusceptometry2021}.
Magnetic ellipsometry expands on this with directionality by measuring the relative amplitudes and phase delays of multiple vector components. 
Existing dc and ac field vector imaging techniques using NV cent\RE{}s \cite{maertzVectorMagneticField2010, steinertHighSensitivityMagnetic2010, ranaImagingMicrowaveMagnetic2025} (or recently, hexagonal boron nitride (hBN) \cite{healeyOmnidirectionalMagneticImaging2026}) can already reconstruct local vector magnetic field magnitudes around such materials, but generally do not exploit the phase relationships between vector field components. 
Because only field amplitudes are measured, intrinsic anisotropy in the permeability tensor can be difficult to distinguish from direction-dependent demagnet\IS{}ation effects arising from sample geometry.

In anisotropic magnetic materials (such as ferrites), an oscillating magnetic field $\mathbf{B}_{\mathrm{AC}}$ induces a secondary magnetic field $\mathbf{B}_{\mathrm{induced}}$ which exhibits direction-dependent phase delays ($\delta$) dictated by the complex permeability tensor $\mu_{ij} = \mu'_{ij} - j\mu''_{ij}$
\cite{laxMicrowaveFerritesFerrimagnetics1962}. 
Here, the real tensor component $\mu'_{ij}$ represents energy storage mechanisms in the material, and the imaginary tensor component $\mu''_{ij}$ represents energy dissipation mechanisms. 
Some of the aforementioned vector imaging studies have already observed anisotropic behavi\OU{}r in such materials using NV cent\RE{}s \cite{grossRealspaceImagingNoncollinear2017}, and hBN \cite{healeyOmnidirectionalMagneticImaging2026}, but this has been limited to dc investigations.
Conversely, by imaging the induced field's 3D polar\IS{}ation ellipse across a material for various excitation angles, the presented method becomes sensitive to the local complex susceptibility tensor of a material.
In this context, the measured field ellipticity plays a role analogous to the polar\IS{}ation observables used in optical ellipsometry \cite{FujiwaraSpectroscopicEllipsometry2007}. 
This could provide information about material parameters such as magnetic anisotropy, permeability, and directional loss tangent [($\tan \delta)_{ii} = \mu''_{ii}/\mu'_{ii}$] which are difficult to obtain through amplitude-only measurements.

To illustrate the simplified 2D process, a Magpylib \cite{ortnerMagpylibFreePython2020} simulation of a 1 mm-radius, 0.5 mm-thick MnZn ferrite cylinder was performed using representative material parameters from a TDK N87 ferrite datasheet \cite{TDKElectronicsFerritesAccessoriesN872025} and from T. Tsutaoka et al.\ \cite{tsutaokaFrequencyDispersionComplex2003}.
The frequency-dependent complex permeability was represented with a simplified model using a multi-term Debye expansion \cite{xuComplexPermittivityPermeability2010, liuBehavioralModelingComplex2021}, and an arbitrary synthetic anisotropy was introduced to provide a clear visual\IS tion response. 
This is not intended to replicate an exact existing ferrite model, but is an order-of-magnitude representation of ferrite behavi\OU r, and gives an indication of how this ellipsometry technique could be used. 
The ferrite model was excited with magnetic field ($\mathbf{B}_{\mathrm{AC}}$) spanning kHz to 10 MHz at a known angle to the preferred material axis, and the resulting induced ($\mathbf{B}_{\mathrm{induced}}$) field was calculated using the susceptibility model and a self-demagnet\IS{}ation estimate. 
Figure \ref{fig:polarimetry-imaging}(b) illustrates one configuration($\mathbf{B}_{\mathrm{AC}}$ at 10 MHz, 45$^\circ$ to the preferred material axis) with greatly exaggerated anisotropy ($\mu_x/\mu_y \approx 5.5$), and the resulting pixel-wise polar\IS{}ation ellipses of the induced field experienced at an NV ensemble above the ferrite.
Figures \ref{fig:polarimetry-imaging}(c, d) show the results of the same analysis performed (with a more feasible anisotropy of $\mu_x/\mu_y \approx 1.15$) on the total field ($\mathbf{B}_{\mathrm{AC}} + \mathbf{B}_{\mathrm{induced}}$) for multiple excitation angles. 
In these figures, realistic anisotropy produces a measurable $\sim 1.5\degree$ change in the average field ellipticity at MHz frequencies compared with an isotropic equivalent sample [Figs.~\ref{fig:polarimetry-imaging}(d) and \ref{fig:polarimetry-imaging}(c), respectively].
Here, the ellipticity was taken as the third Stokes parameter $S_3$ \cite[p.~12]{CollettObservablesPolarizedLight2005} defined by:
\begin{align} \begin{aligned}\chi_{\mathrm{ellipse}} &=
	\frac{1}{2} \arcsin \left( 	\frac{S_3}{S_0} \right) \\
	&= \frac{1}{2} \arcsin\left( 
		\frac{ 
			2\left|B_x\right|\left|B_y\right| 
			\sin\left(\arg\left(B_y, B_x^{*}\right) \right) 
		}
		{ \left|B_x\right|^2+\left|B_y\right|^2 } 
		\right). \label{eq:ellipticity} 
\end{aligned} \end{align}

With the sample geometry and excitation field known, the complex 3D susceptibility tensor could in principle be inferred by optim\IS{}ing a forward model until its predicted field-polar\IS{}ation ellipses match measurement values. 
Access to out-of-phase induced field components provides additional independent information, improving the uniqueness and conditioning of complex susceptibility tensor reconstruction relative to amplitude-only measurements.
Additionally, sweeping the excitation direction reveals the angular symmetries of the response, and the points of zero ellipticity, providing initial estimates of the principal-axis orientations of an anisotropic material.
Extending the excitation sweep into three dimensions provides sensitivity to all tensor components.
Solving Eq.~\ref{eq:amplitude-fitting} for measured polar\IS{}ation ellipses, and repeating the susceptibility optim\IS{}ation at each frequency reconstructs the full frequency-dependent susceptibility tensor $\mathrm{\chi_{\textit{ij}}(\omega)}$.
The achieved phase uncertainty in the above crossed-coil experiments suggests sensitivity to material loss tangents on the order of $\tan \delta \approx 0.01$ resolvable on the micromet\RE{} scale, although validation on real materials remains future work.

\section{CONCLUSION} 
We have presented a technique for measuring the polar\IS{}ation of ac magnetic fields in three dimensions, by extending existing phase-sensitive NV-ODMR vector magnetometry methods, without requiring real-time baseband readout. 
The method reconstructs the complex three-dimensional magnetic field vector and its polar\IS{}ation ellipse from the relative amplitudes and phases of the NV spin-resonance responses.

We demonstrated the technique through two proof-of-concept measurements. 
Rotation of a single source coil produced the expected rotation of the reconstructed magnetic field vector, while also revealing weak ellipticity associated with eddy currents induced in a nearby optics table. 
Measurements using two orthogonal coils demonstrated sensitivity to elliptical magnetic field polar\IS{}ations, and identified out-of-phase field components arising from mutual inductive coupling between the coils.

Finally, simulations of an anisotropic MnZn ferrite demonstrated how spatially resolved magnetic field polar\IS{}ation measurements could provide sensitivity to directional permeability and magnetic loss tangent. These results establish magnetic polarimetry as a practical extension of ac vector magnetometry and motivate its application to coupled electrical systems and anisotropic magnetic materials.

\textit{Acknowled\GE{}ments.}
MSJB acknowledges support from the Office of National Intelligence's (ONI) National Intelligence Postdoctoral Grant (NIPG) 2022 and the Australian Army's Robotic and Autonomous Systems Implementation \& Coordination Office's (RICO) Quantum Technology Challenge (QTC) 2022.
We would like to thank Alex Tritt for helpful feedback on the manuscript.

\textit{Author contributions.}
M.S.J.B. conceived the idea, performed the experiments and co-wrote the manuscript.
T.J.C. assisted with the experiments. 
J.P.D. conducted with ferrite simulations, assisted with data analysis, and co-wrote the manuscript.
K.H. provided supervision and feedback on the manuscript.

\textit{Data availability.}
The data that support the findings of this study are available from the corresponding author upon reasonable request.

\clearpage

\section{APPENDIX - METHODS} 
For the rotating coil experiment, a single coil was placed approximately 30 cm away from the diamond and rotated in the horizontal plane. 
The coil was driven by a \textsc{SR860} LIA at 777 Hz and a \textit{Crown Audio} \textsc{XLS 1002} audio power amplifier. 
The microwave generator was a \textit{Marconi} \textsc{2032 microwave generator} which was also used for FM of the microwaves. 
Microwaves were amplified using a \textit{Mini-Circuits} \textsc{ZHL-16W-43-S+} amplifier. Microwaves were delivered to the sample using a monolithic loop-gap resonator co-axially excited from a single-turn antenna. 
The green optical excitation was provided by a \textsc{PT54} LED and red PL measured using a custom photodiode and transimpedance amplifier assembly, with about 10 MHz bandwidth. 
Both the PL and excitation light were directed to the diamond using tapered light-pipes (\textit{Edmund Optics}, \textsc{\#63-103}). 
The green excitation was optically filtered at the photodiode. 
The diamond was an \textit{Element6} \textsc{DNVB14}. 
ODMR sweeps took 10 seconds to sweep from 2.7 to 3.1 GHz and the LIA amplifier time-constant was set to 3 ms. 

For the crossed coil experiment, the coils were placed orthogonal to each other and driven by the L+R channels of the same audio power amplifier.
One channel was phase shifted by $\pi/2$ using a simple transistor-based phase shifting circuit, the ac field was at a frequency of 1.77 kHz.
The coils were both approximately 10 cm away from the diamond and the diamond was placed on a microwave PCB resonator which was suspended midair by stainless steel mounts.
The microwave generator, amplifier and diamond were the same as for the rotating coil experiment.
For this experiment a Zurich Instruments H2FLI lock-in amplifier was used with the HF2TA current amplifier amplifying the photocurrent from a Thorlabs DET36A2 photodiode.
The ODMR sweeps from 2.5 to 3.2 GHz took 20 seconds and the LIA time constant was set to 47.4 ms.
Green excitation light came from a Laser Quantum GEM 532 nm laser with a power of about 100 mW into a dichroic and a 0.8 NA objective lens.

\bibliography{polarisation-paper.bib}

\end{document}